\documentclass[aps,prl,reprint,eprint,longbibliography]{revtex4-1}
\eprint
\usepackage{graphicx}
\usepackage{dcolumn}
\usepackage{bm}
\usepackage[utf8]{inputenc}
\usepackage{color}
\usepackage{amsmath}
\usepackage{amsfonts}
\usepackage{graphicx}
\usepackage{color}

\usepackage{amssymb}
\usepackage{wasysym}

\begin{document}
\title{Photon transport through the entire adult human head}
\author{Jack Radford, Vytautas Gradauskas, Kevin J. Mitchell, Samuel Nerenberg, Ilya Starshynov, and Daniele Faccio}
\email{daniele.faccio@glasgow.ac.uk}
\affiliation{School of Physics and Astronomy, University of Glasgow, Glasgow, G12 8QQ, UK}

\begin{abstract}
Optical brain imaging technologies are promising due to their relatively high temporal resolution, portability and cost-effectiveness.  
However, the highly scattering nature of near-infrared light in human tissue makes it challenging to collect photons emerging from more than 4 cm below the scalp, or with source-detector separation larger than several centimeters. We explore the physical limits of photon transport in the head and show that despite an extreme attenuation of $\sim10^{18}$, we can experimentally detect light that is transmitted diametrically through the entire adult human head. Analysis of various photon migration pathways through the head also indicates how the source-detector configuration can be used to isolate photons interacting with deep regions of the brain that are inaccessible with current optical techniques.
\end{abstract}

\maketitle
\begin{figure}[t!]
 \centering
 \includegraphics[width=1\linewidth]{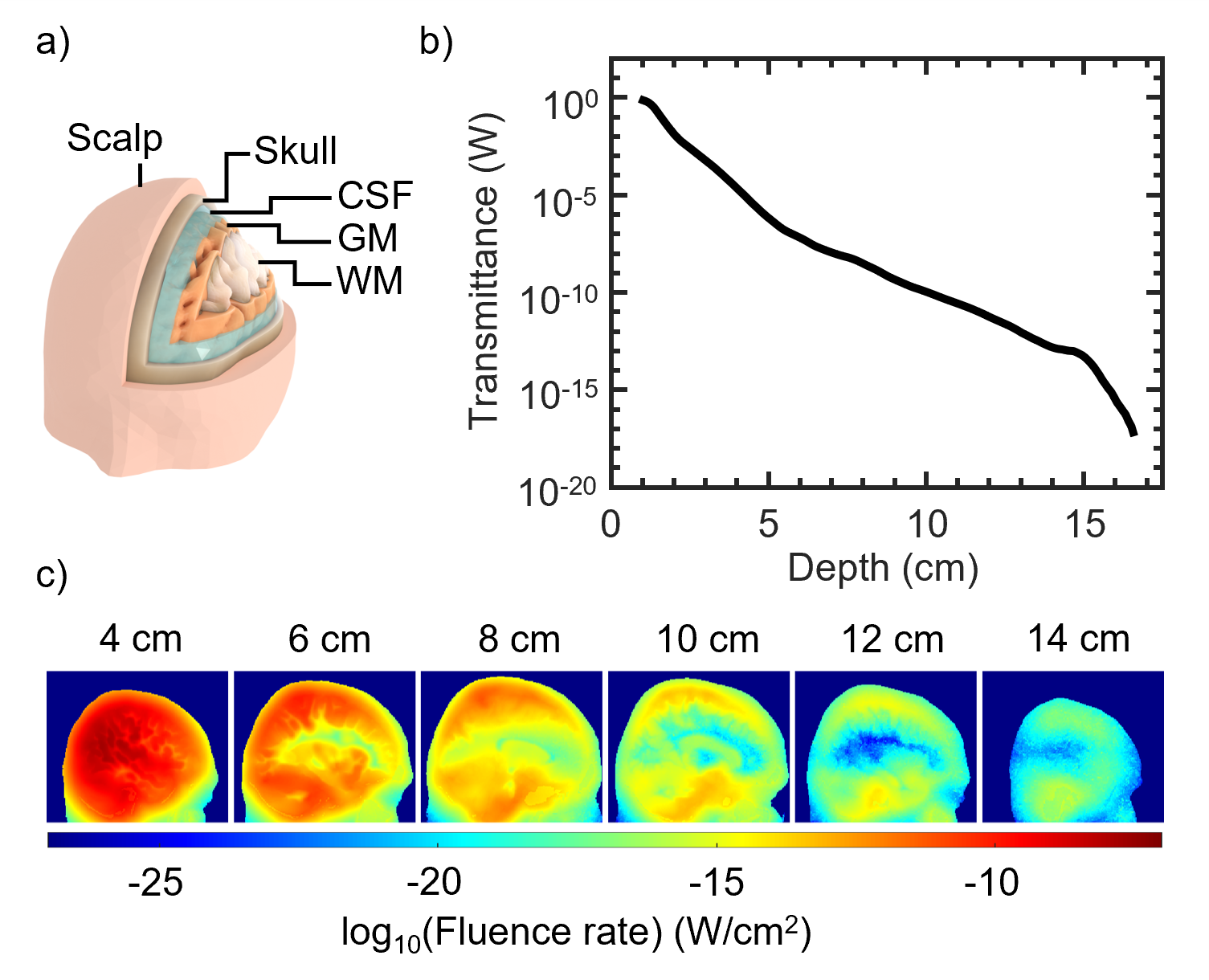}
 \caption{Numerical simulations. a) A smoothed render of the head volume mesh used in simulations highlighting the 5 layers considered in this work. b) Numerically estimated transmittance of light, diametrically through the head for a 1W source. c) The distribution of fluence rate (W/cm$^2$) at various sagittal slices plotted in logarithmic scale - light preferentially persists in regions of low absorption and scattering.
 }
\label{fig:attenuation}
\end{figure}
\begin{figure}[t]
 \centering
 \includegraphics[width=1\linewidth]{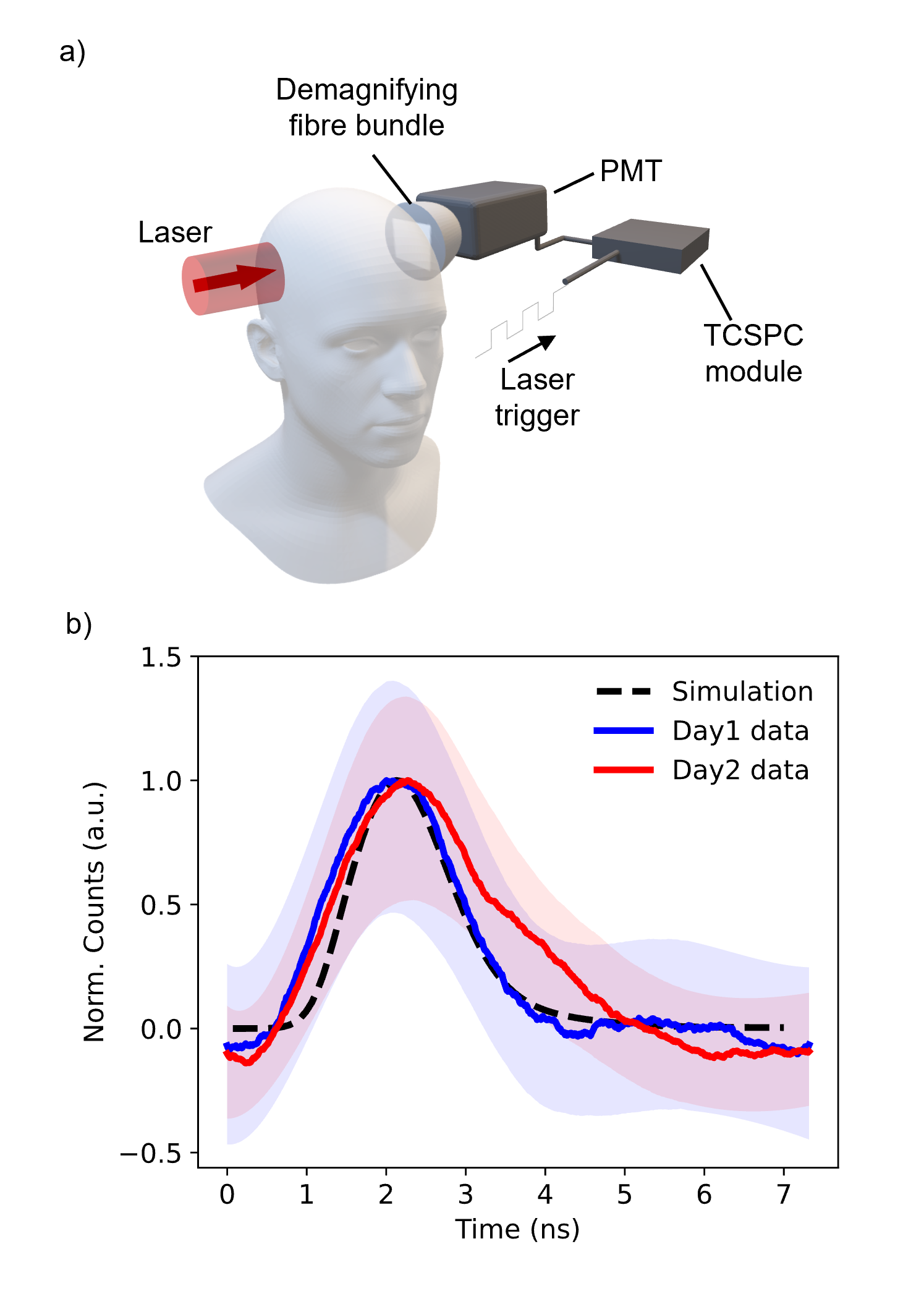}
 \caption{Experimental results. a) The experimental configuration used in laboratory experiments - an ultrafast pulsed laser (1.2 W power, 800 nm wavelength, 140 fs pulse duration, 80 MHz repetition rate) is expanded to a uniformly distributed 25 mm diameter circle and projected against the side of the head. Diametrically opposite the source, a photomultiplier tube (PMT) is synchronised to the laser trigger such that a histogram of photon time-of-flight (ToF) can be measured using TCSPC. b) The normalised simulated ToF (black dashed) and the mean of 15 experimental trials (2 minute exposure) on day 1 (blue), and day 2 (red). The uncertainty bands indicate the standard error of the mean at each time stamp. }
\label{fig:tof}
\end{figure}

{\bf{Introduction.}}
Optical modalities for non-invasive imaging of the human brain hold promise to fill the technology gap between cheap and portable devices such as electroencephalography (EEG) and expensive high resolution instruments such as functional magnetic resonance imaging (fMRI), with devices that have high sample rates, high spatial resolution, and are relatively inexpensive. A major bottleneck for the widespread adoption of optical brain reading devices in clinics and neuroscience studies is the low number of photons emerging from deep layers of the brain that restrict the sensitivity of these methods to a maximum of 4 cm below the scalp \cite{Strangman2013DepthTemplate, Mora2015TowardsSensitivity, Dehghani2009_DepthTomography}, corresponding to the outermost layer of the cerebral cortex (gyri). However, imaging paradigms have been demonstrated using highly scattered transmitted photons in thick diffusive materials \cite{Lyons2019, Radford2020RoleImaging} and it has been theoretically shown that photons experiencing similar scattering length scales as the diameter of the human head can carry imaging information \cite{Radford2023InformationRegime}. Therefore, extending current optical methods to extract information about deep brain regions that are currently inaccessible, e.g., cortical folds (sulci), midbrain, and deep regions of the cerebellum, may be feasible with careful consideration of sources, detectors and incorporating time-resolved photon counting information so long as photons are detected through the extreme number of attenuation lengths. \\
The most common optical approach to infer brain activity measures differential changes in absorption of near-infrared light to determine blood oxygenation levels. One of the pioneering works to demonstrate this technique was reported by Jobsis \cite{Jobsis1977} and later led to the field of functional near-infrared spectroscopy (fNIRS). Current fNIRS devices have faster sample rates than functional magnetic resonance imaging (fMRI) and better spatial resolution than electroencephalograms (EEG) \cite{Pifferi2016, Scarapicchia2017_fMRI_fNIRS}. Diffuse optical tomography (DOT) extends fNIRS to include depth information which can be used to selectively resolve changes in blood chromophore in superficial and cortical layers \cite{Saager2008MeasurementSpectroscopy, NicholasM.Gregg2010BrainTomography, Lee2017, Wheelock2019}. \\
Unfortunately, there remain outstanding challenges associated with the use of optical and near-infrared wavelengths in the adult human head, the most prominent of which is the shallow depth of sensitivity under the scalp that restricts fNIRS to monitor only the outer-most region of the cortex. This limitation is due to the highly scattering nature of human tissue which causes an exponential attenuation of light for increasing penetration depth. 
Liu et al. attempted to measure deep brain activity with fNIRS  \cite{Liu2015InferringSpectroscopy} using a model which maps co-registered fNIRS to fMRI data and can be used to infer activity in deep regions (sulci) from surface-level (gyri) optical measurements. However, a direct measurement of this information is more accurate than relying on correlations between fNIRS and fMRI, which monitor brain activity by fundamentally different physical mechanisms. \\
In the claims of the early work by Jobsis in \cite{Jobsis1977}, a signal is presented highlighting an increase in transmission of near-IR light diametrically from temple to temple in an adult human head due to a decrease in cerebral blood volume during hyperventilation. Unfortunately, the results were incomplete due to stopping the experiment before the signal could return to baseline. Since this work, the only studies to detect light diametrically (i.e. across the widest point of the skull) through the head involve neonatal or infant subjects that have more transparent skulls and significantly smaller diameter heads compared with adults. Studies in the infant brain has shown that fNIRS allows for 3D tomographic reconstructions of brain activity and cerebral haemodynamics of the entire head \cite{Lee2017,Hebden2002Three-dimensionalBrain,Benaron2000NoninvasiveLight}. However, using transmitted light to produce tomographic maps in this way for the adult head has yet to be demonstrated. Indeed, some suggest that the detection of light diametrically through an adult human head is ``impossible'' \cite{Becker2005}. Indeed, a back-of-the-envelope estimate that very simply considers only the white matter region ($L=10$ cm thick, $\mu_a=0.92$cm$^{-1}$, $\mu_s'=49.4$cm$^{-1}$ \cite{Cassano2019SelectiveStudy}) and an exponential Beer-Lambert decay, $\sim\exp{(-\sqrt{3\mu_a(\mu_a+\mu_s'})L}$ \cite{Yaroshevsky2011} would indicate an remarkably high attenuation of $10^{53}$ thus supporting the apparent impossibility of detecting photons that have propagated diametrically through the brain. \\ 
In this work, we explore the details of photon transport through the head 
and present numerical and experimental evidence that photons can actually be transmitted diametrically through the entire adult human head, albeit along specific trajectories. We find a broad variety of light propagation pathways that are largely determined by the presence of and guiding by cerebrospinal fluid. Different source positions on the head can then selectively isolate and probe deep regions of the brain. This suggests optical techniques could be used to monitor activity in the sulci, midbrain, and deep regions of the cerebellum, which are currently inaccessible with fNIRS \cite{Chen2020fNIRSClinical}. These findings uncover the potential to extend non-invasive light-based brain imaging technologies to tomography of critical biomarkers deep in the adult human head.\\
%
%
\textbf{Numerical modelling.}
Monte Carlo ray-tracing simulations were performed using Monte Carlo eXtreme (MCX) to estimate the attenuation and distribution of light through the head. An open source five-layer head volume mesh (Fig.~\ref{fig:attenuation}a) derived from an averaged MRI \cite{Tran2020ImprovingModels} was used with optical coefficients from \cite{Cassano2019SelectiveStudy} (at 810nm). The estimated attenuation of light diametrically through the head is of the order $10^{18} - 10^{20}$ (Fig.~\ref{fig:attenuation}b). To overcome this attenuation, $7.24 \times 10^{13}$ single-point precision photon packets were used, each with an initial weight of 1. The simulation used in Fig.~\ref{fig:attenuation} required $>$850 hours run-time with an NVIDIA GeForce RTX 4090 graphics card.\\
%
Analysis of the simulated fluence-rate distribution through the head in Fig.~\ref{fig:attenuation}c) shows that light explores all regions of the brain and persists mostly in areas with low scattering and absorption, such as the cerebrospinal fluid above and below the cortex. \\ 
\textbf{Experimental evidence.}
To demonstrate the feasibility to detect photons that have experienced the longest trajectories, and therefore most likely to interact with the deepest layers of the head, an experiment was performed to detect light that has travelled diametrically from one side of the head to the other (i.e. across the widest point of the skull). \\
The experimental configuration is outlined in Fig.~\ref{fig:tof}a):  a pulsed laser (1.2 W power, 800 nm wavelength, 140 fs pulse duration, 80 MHz repetition rate) is expanded to a uniformly distributed circular beam of 25 mm diameter and projected against the side of the head above the ear. Diametrically opposite the source, a demagnifying tapered fiber bundle (Edmund Optics 25 mm to 8 mm Fiber Optic Taper) is placed in close proximity to the scalp and redirects light to a photomultiplier tube (PMT, Hamamatsu H7422P-50). The PMT operates in photon counting mode, such that the detection of a photon produces an electrical pulse that can be synchronised with the laser emission to produce a photon time-of-flight (ToF) distribution using a time-correlated single-photon counting (TCSPC) module (Becker \& Hickl SPC-150N). \\
The combination of high laser power and a large area single-photon sensitive detector is optimised to overcome the extreme attenuation of light through the head. Increasing the area of incident laser exposure allows the use of high powers without reaching the maximum permissible exposure of human skin. This is in contrast to typical fNIRS devices that use point-like emitters in an effort to maintain high spatial resolution, as well as low-power and small form-factor hardware. \\ 
The detector used in this work is chosen to maximise the etendue (product of light collection solid angle and sensor area) of light collection with a large area detector (5 mm diameter). The demagnifying tapered fiber bundle is used to further increase the area of collection by $\sim5\times$ whilst maintaining a high numerical aperture (NA=1). This differs from typical fNIRS devices which restrict the area and solid angle of collection of the detectors with light-guides or conventional optical fiber bundles with comparatively small collection area and numerical aperture. Finally, the PMT used in this work has a very low dark count rate (15 cps), is single-photon sensitive (QE=15\% at 800 nm), and has a low timing jitter (IRF FWHM=300 ps) such that the transmitted photons can be correlated with the source using TCSPC to overcome background photon counts and noise.\\
The results in Fig.~\ref{fig:tof}b) show experimental ToF distributions for photons transmitted through an adult male head (15.5 cm diameter) collected on two different days, 1 week apart. For each day, time traces are the mean of 15 2-minute exposures, resulting in 30 minutes of total acquisition time. The uncertainty bounds of the time traces represent the standard error of the mean at each time stamp. Both the data and the uncertainty bounds were smoothed using a Savitsky-Golay filter. The measured experimental attenuation was found to be of the order $10^{18}$, corresponding to a detection of around 1 photon per second for a 1.2 W source. The simulated ToF distribution for the experimental configuration was approximated using the detected photons diametrically opposite the source using kernel density estimation with the kernel bandwidths determined by Silverman’s rule \cite{Silverman1998} to overcome sub-sampling. The distribution is then convolved with the experimental impulse response function. The simulated result shows good agreement with experimental results for the first two moments of the distribution, i.e. for both the peak delay time and width of the ToF distribution, which indicates that the measured signal propagated through the head via similar migration pathways as seen in the numerical model.\\
\textbf{Photon migration pathways.}
Despite considering photon propagation in the human head as a diffuse optics problem where it is typically expected that light is highly scattered in all directions, the heterogeneity of optical properties and complex geometry of different layers causes light to be guided through the head in preferred pathways. This phenomena is caused by channels of low scattering and absorption (e.g., cerebrospinal fluid) surrounded by high scattering (e.g., skull and grey matter), such that light follows the path of least extinction. A general study of the fundamental mechanism of guided light in diffusive materials was recently explored in \cite{Mitchell2024}. 
\begin{figure}[t!]
 \centering
 \includegraphics[width=\linewidth]{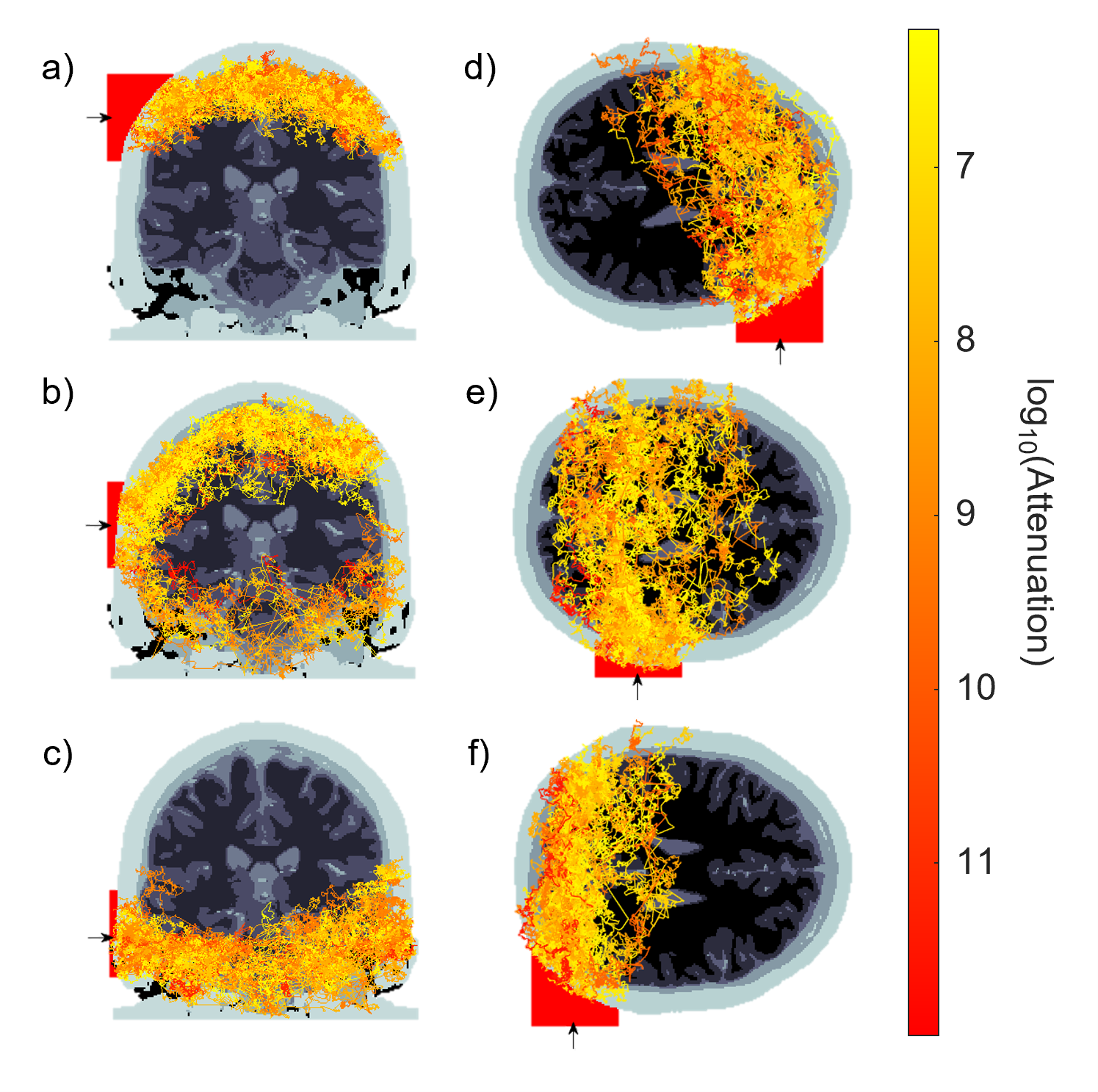}
 \caption{Numerical simulations: a 2-dimensional projection of 50 random transmitted photon paths when moving the source, represented by an arrow and red area (5 cm diameter) outside of the head from the top to the bottom (a to c) and from the front to the back of the head (d to f). The positions in (b) and (e) are approximately the positions of the source and detector in experimental measurements. 
 The detector position and area remains fixed in all cases.
 }
\label{fig:trajectories}
\end{figure}
In the context of fNIRS, light guided by the cerebrospinal fluid has been observed in numerous previous studies and is typically regarded as an unwanted nuisance that decreases the depth of penetration of light and increases uncertainty when attempting to localise brain activity \cite{Okada2003_CSFmodeling, Wolf1999ThePhantom, Custo2006EffectiveImaging}. By contrast, in this work we argue that the role of light guided by weakly scattering channels is critical for localising changes in light absorption in deep brain regions and can be exploited by optimised source and detector arrangements.  \\
Analysis of 50 random detected photon packet trajectories for various source positions is shown in Fig.~\ref{fig:trajectories}. Positioning the source high above the ear can cause light to be guided around the top of the brain in the CSF layer (Fig.~\ref{fig:trajectories}a), whereas lower positions cause light to be transported under the brain (Fig.~\ref{fig:trajectories}c).
Likewise, moving the source towards the back of the head or forwards towards the temple can increase the likelihood of photon pathways reaching the occipital or frontal lobe regions, respectively (Fig.~\ref{fig:trajectories}d-e). \\
The position of the source in Fig.~\ref{fig:trajectories}b) and e) closely approximates the configuration used in the experiment. A source launch area of 5 cm diameter was used to account for the uncertainty in position throughout the experiments. Surprisingly, although it is likely that most of the experimentally detected photons propagated around the top of the head, Fig.~\ref{fig:trajectories}b) indicates that there are also trajectories under the cerebrum with similar order of magnitude attenuation. \\
\begin{figure}[t!]
 \centering
 \includegraphics[width=1\linewidth]{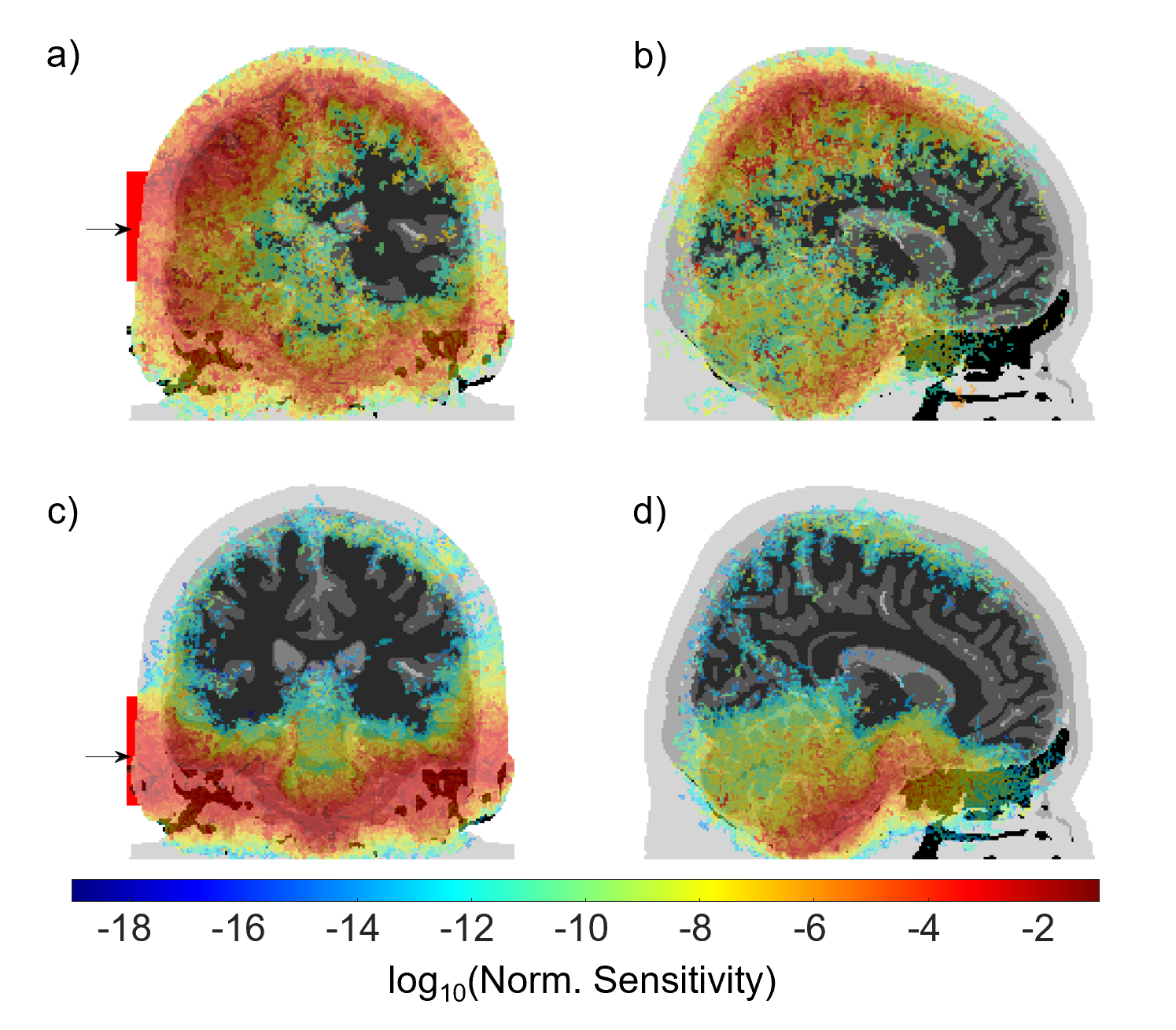}
 \caption{The sensitivity map (Jacobian matrix) for the source position  approximately matching the experimental conditions (a and b) and  40 mm lower than experimental conditions (c and d).
 }
\label{fig:sensitivity}
\end{figure}
We also show in Fig.~\ref{fig:sensitivity}, a sensitivity analysis that is obtained by calculating the Jacobian for the light rays, i.e. a map of how light intensity changes at the detector for small changes to the absorption coefficient at each voxel in the head \cite{Yao2018Directreplay}. 
The Jacobian for the source-detector arrangement approximating experimental conditions, Fig.~\ref{fig:sensitivity}a) and b), highlights that measurements in this configuration are indeed sensitive to changes in absorption in deep regions of the brain reaching areas that are currently inaccessible for fNIRS devices, such as the midbrain, sulci, and deep regions of the cerebellum. Figure~\ref{fig:sensitivity}c) shows that by repositioning the source 40 mm lower, isolates the sensitivity map almost exclusively to regions under the cerebrum and demonstrates the potential to target deep brain regions with careful consideration of source-detector arrangements. These Jacobians are an extension of the typical ``banana-like" sensitivity profiles found in fNIRS \cite{Strangman2013DepthTemplate, Cui1991StudySpectroscopy} in the limit of large source-detector separation which here evolve into non-trivial shapes that could be used to reconstruct tomographic information.\\
\textbf{Conclusions.}
Detecting photons transported through large and diametrically opposite source-detector separations has the potential to extend the field of optical brain imaging devices to reach regions of the brain currently considered inaccessible. Experimental measurements of photons transmitted through the entire adult human head suggest that, although the attenuation of light is challenging, in principle it is possible to detect light from the most extreme source-detector separations. \\
Although we find that the numerical simulation of the experimental measurement has relatively close qualitative agreement to the experimental data, we expect discrepancies given the large uncertainty of optical properties of the head layers \textit{in vivo} which can vary by 100\% in the literature \cite{Jacques2013, Farina2015In-vivoHead}. We also expect deviations between the simulated and experimental data due to the differences in structure, shape and thickness of the layers between the open-source mesh used in simulations versus the participant's anatomy. Although this will almost certainly cause the simulated photon migration pathways to differ from reality in the fine details, we expect the observed guiding principle e.g., above and below the cerebrum, to remain accurate. \\
Furthermore, careful consideration of source-detector configuration and time-of-flight analysis present an opportunity to selectively isolate photons that are confined to guided propagation pathways that could possibly be combined to reconstruct tomographic information of deep brain activity. 

{\bf{Acknowledgements.}} The authors acknowledge funding from EPSRC (UK, grants no. EP/T00097X/1, EP/T021020/1, EP/Y029097/1). DF is supported by the Royal Academy of Engineering under the Chairs in Emerging Technologies scheme. \\

{\bf{Declarations.}} Participants involved in this work provided informed consent before experiments and the experiments were approved by the University of Glasgow College of Science and Engineering Ethics Committee (application number: 300180292).\\

{\bf{Author Contributions.}} CRediT: JR: conceptualisation, conceptualisation, methodology, investigation, formal analysis, data curation, visualization, supervision, writing - original draft. VG: investigation, visualisation, writing - review \& editing. KM: investigation, visualisation, writing - review \& editing. SN: investigation, methodology, supervision,  writing - review \& editing. IS: investigation, writing - review \& editing. DF: conceptualisation, methodology, writing - original draft, supervision, project administration, funding acquisition.\\

{\bf{Disclosures.}} The authors declare no conflicts of interest. \\

\bibliographystyle{ieeetr}
\bibliography{references} 

\end{document}